\documentclass[11pt,twoside,letterpaper]{article} 
\usepackage{times,fancyhdr}
\usepackage[dvips]{graphicx}
\usepackage{graphicx}
\usepackage{dcolumn}
\usepackage{mathrsfs}
\usepackage{amsmath}
\usepackage{bm}
\usepackage{amssymb} 
\makeatletter 
\def\cleardoublepage{\clearpage\if@twoside \ifodd\c@page\else%
    \hbox{}%
    \thispagestyle{empty}%
    \newpage%
    \if@twocolumn\hbox{}\newpage\fi\fi\fi} 
\makeatother 
\renewcommand{\thesection}{\arabic{section}.}
\renewcommand{\thesubsection}{\thesection\arabic{subsection}.}

\def\figurename{Figure}
\makeatletter
\renewcommand{\fnum@figure}[1]{\figurename~\thefigure.}
\makeatother
\def\tablename{Table}
\makeatletter
\renewcommand{\fnum@table}[1]{\tablename~\thetable.}
\makeatother
\begin{document}
\title{ 
{\begin{flushleft}
\vskip 0.45in
{\normalsize\bfseries\textit{Chapter~1}}
\end{flushleft}
\vskip 0.45in
%
%
%
%
\bfseries\scshape Influence of the nuclear
symmetry energy on the structure and composition of the outer crust}}
\author{\bfseries\itshape X. Roca-Maza\thanks{E-mail: xavier.roca.maza@mi.infn.it}\\ 
INFN, Sezione di Milano, 20133 Milano, Italy.\\ 
\bfseries\itshape J. Piekarewicz\thanks{E-mail: jpiekarewicz@fsu.edu} \\ 
Department of Physics, Florida State University, Tallahassee, FL 32306, USA.\\
\bfseries\itshape T. Garc\'ia-G\'alvez\thanks{E-mail: mtgcia@gmail.com} 
\ and M. Centelles\thanks{E-mail: mariocentelles@ub.edu}\\
Departament d'Estructura i Constituents de la Mat\`eria and\\ Institut de 
Ci\`encies del Cosmos, Facultat de F\'{\i}sica, \\Universitat de Barcelona,
Diagonal {\sl 647}, {\sl 08028} Barcelona, Spain.}
\date{}
\maketitle
\thispagestyle{empty}
\setcounter{page}{1}
\thispagestyle{fancy}
\fancyhead{}
\fancyhead[L]{In: Neutron Star Crust \\ 
Editors: C.A. Bertulani and J. Piekarewicz, pp. {\thepage-\pageref{lastpage-01}}} 
\fancyhead[R]{ISBN 0000000000  \\
\copyright~2012 Nova Science Publishers, Inc.}
\fancyfoot{}
\renewcommand{\headrulewidth}{0pt}
\vspace{0.2in}

\pagestyle{fancy}
\fancyhead{}
\fancyhead[EC]{X. Roca-Maza, J. Piekarewicz, T. Garc\'ia-G\'alvez, and M. Centelles}
\fancyhead[EL,OR]{\thepage}
\fancyhead[OC]{Influence of the symmetry energy on the outer crust}
\fancyfoot{}
\renewcommand\headrulewidth{0.5pt} 
%
%
\begin{abstract}
We review and extend with nonrelativistic nuclear mean field calculations a 
previous study of the impact of the nuclear symmetry energy on the structure and
composition of the outer crust of nonaccreting neutron stars \cite{Roca-Maza:2008}.
First, we develop a simple {\sl ``toy model''} to understand the most
relevant quantities determining the structure and composition of the
outer crust: the nuclear symmetry energy and the pressure of the
electron gas. While the latter is a well determined quantity, the
former ---specially its density dependence--- still lacks an
accurate characterization. We thus focus on the influence of the nuclear 
symmetry energy on the crustal composition. For that, we employ different  
nuclear models that are accurate in the description of terrestrial nuclei.
We show that those models with stiffer symmetry energies ---namely,
those that generate thicker neutron skins in heavy nuclei and have
{\sl smaller} symmetry energies at subnormal nuclear
densities--- generate more exotic isotopes in the stellar crust than
their softer counterparts.
\end{abstract}
%
%
\section{Introduction}
\label{Introduction}

Massive enough stars at the final stage of their evolution, produce the
so-called supernovae explosion whose remnant is a neutron star or a
black hole. Because all the nuclear fuel has been exhausted in fusion
reactions during the lifetime of the progenitor star, the neutron star
cannot compensate the gravitational pressure by burning light elements
into heavier ones. Therefore, as there is no nuclear reaction
equilibrating the gravitational pressure, the stability has to be
ensured via the stiffness of the nuclear strong interaction when the
baryonic matter is compressed to extreme densities by gravity. This
new type of equilibrium leaves formidable imprints on the neutron star
as compared to an average star. For instance, a typical neutron star
with a mass like our Sun has a radius of about only $10$ {\rm km},
{\it i.e.}, almost five orders of magnitude smaller than the radius of
the Sun. Table~\ref{nsprop} collects some 
typical values for a few properties of a neutron star. Owing to the
huge densities and pressures present in these neutron-rich celestial
bodies, the study of neutron stars requires a deep knowledge of the
structure of matter under extreme conditions of density and isospin
asymmetry~\cite{Lattimer:2000nx,Lattimer:2004pg}.

\begin{table}[b]
\begin{center}
\begin{tabular}{|c|c|c|c|c|}
\hline
$R$ (km)&$\bar{\rho}$ (gr/cm$^3$)&$v_{\rm esc}/c$&$g/g_{\rm Earth}$ (surface)&$P$ (dyn/cm$^2$)\\
\hline
      10&$10^{14} - 10^{15}$      &            0.5&                $10^{11}$ & $0 - 10^{35}$   \\
\hline
\end{tabular}
\caption{
    Properties of a typical 1 solar-mass neutron star.
    Here $R$ is the radius, $\bar{\rho}$ is the mean density, 
    $v_{\rm esc}$
    is the escape velocity, $c$ is the speed of light, $g$ is the
    acceleration of gravity and $P$ is the pressure.}   
\label{nsprop}
\end{center}
\end{table}

\begin{figure}[t]
 \centerline{ \includegraphics[clip=true, width=0.6\linewidth]{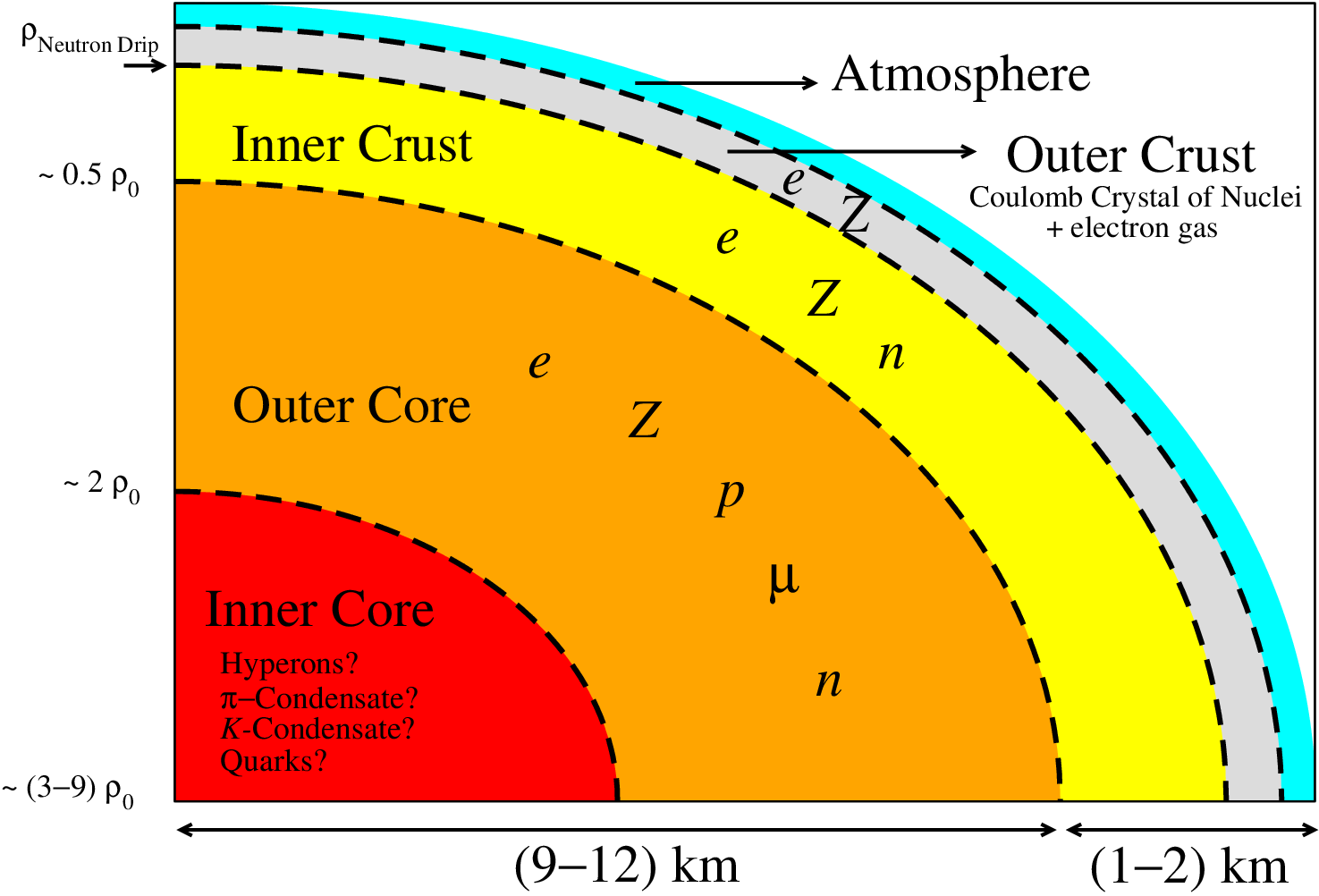}}
 \vspace{-0.2cm}
 \caption{Rendition of the assumed structure and composition of
           a neutron star.}
 \label{Fig01}
\end{figure}

Figure~\ref{Fig01} is believed to represent a plausible rendition of the structure and 
composition of a neutron star. The outer crust is organized into a Coulomb lattice of 
neutron-rich nuclei embedded in a uniform electron gas~\cite{Baym:1971pw}.  As the 
density increases, nuclei become more neutron rich until the neutron drip region is 
reached. This density defines the outermost part of the inner crust. The inner crust 
also consists of a Coulomb lattice of neutron-rich nuclei, but now, embedded in a 
uniform electron gas and a neutron vapor that permeates the system. As the density 
continues to increase in the inner crust, the system is speculated to morph into a variety
of complex and exotic structures, such as spheres, cylinders, rods,
plates, {\it etc.} ---collectively known as {\emph{nuclear
pasta}}~\cite{Ravenhall:1983uh,Hashimoto:1984,Lorenz:1992zz}. As the 
density increases even further, uniformity is eventually restored.  Finally,
at ultra high densities it has been established that the ground state
of hadronic matter becomes a color superconductor in a
color-flavor-locked (CFL) phase~\cite{Alford:1998mk,Rajagopal:2000wf}. 
It is unknown, however, if the density at the core of a neutron star may 
reach the extreme values required for the CFL phase to develop. Thus, other 
exotic phases ---such as meson condensates, hyperonic matter, and/or quark
matter--- may be likely to harbor the core of neutron
stars. 

Observers have recently discovered high-frequency oscillations in the tails of giant  
flares~\cite{Thompson:1995gw,Kouveliotou:1998ze, Kouveliotou:2003tb} which
are believed to be associated with seismic vibrations of the neutron star
crust~\cite{Piro:2005jf,Watts:2006mr}.
The frequencies of such vibrations are known to be very sensitive to
the composition of the crust and to the nuclear symmetry energy and its density
dependence~\cite{Strohmayer:2006py,Watts:2006ew,SteinerWatts}. 
Moreover, oscillating neutron stars are likely emitters of
gravitational waves. It has been shown that the density dependence of
the nuclear symmetry energy also has a notable influence on the
emission of gravitational radiation from axial modes of
oscillating neutron stars~\cite{GRAVIT}.
This fact may have observational consequences, as it is expected that
gravitational waves emitted by neutron stars can be detected by the
Laser Interferometer Gravitational Wave Observatory (LIGO) and the
future Laser Interferometer Space Antenna (LISA) mission.

The nuclear symmetry energy $E_{\rm sym}(\rho)$ represents the energy
cost involved in converting the protons into neutrons in symmetric
nuclear matter at density $\rho$. This quantity entails important
consequences for a myriad of properties of neutron-rich matter both on
Earth and in the Cosmos. Actually, neutron stars can synthesize very
neutron-rich nuclei, far more isospin asymmetric than we can produce
in terrestrial laboratory conditions. All in all, neutron stars are a
privileged source for observational studies that may help in
constraining nuclear properties under very extreme conditions. We
expect that as the observational techniques improve, neutron stars
will provide stringent limits on the equation of state of neutron-rich
matter. Motivated by this possibility, we study the sensitivity of the
structure and composition of the outermost layer ---apart from the
``atmosphere'', see Fig.~\ref{Fig01}--- of a nonaccreting neutron star
to the model dependence of the nuclear symmetry energy.

The outer crust comprises about seven orders of magnitude
in density; from about $10^{4}{\rm g/cm^{3}}$ up to a neutron-drip density of about
$4\times10^{11}{\rm g/cm^{3}}$~\cite{Baym:1971pw}. At these densities the electrons 
---present to maintain charge neutrality--- are no longer bound to nuclei and move 
freely throughout the crust. Moreover, at these low nuclear densities it is
energetically favorable for the nuclei to arrange themselves in a crystalline lattice. 
At the lowest densities, the electronic contribution is negligible so the Coulomb 
lattice is populated in good approximation by ${}^{56}$Fe nuclei. However, as the 
density increases and the electronic contribution becomes important, the more 
energetically advantageous process for the system is to lower its electron fraction 
by allowing the energetic electrons be captured by protons. The excess energy is 
carried away by neutrinos. The resulting nuclear lattice is then formed by nuclei 
having a slightly lower proton fraction than ${}^{56}$Fe. As the density continues to 
increase, the nuclear system evolves into a Coulomb lattice of progressively more 
neutron-rich nuclei until the critical neutron-drip density is reached. The essential 
physics of the outer crust is then nicely captured by a competition between the 
electronic contribution, which favors the formation of neutron-rich nuclei, and the
nuclear symmetry energy, which favors a more symmetric configuration of the lattice nuclei. 

It is important to notice that the neutron-rich nuclei populating the Coulomb lattice in the 
outer crust are on average more dilute than their more symmetric stable counterparts because 
of the development of a neutron skin. Let us recall that the size of the
neutron skin of a nucleus is usually characterized by the so-called
neutron skin thickness
\begin{equation}
\Delta r_{np} = r_n-r_p \ ,
 \label{rnp}
\end{equation}
where $r_n$ and $r_p$ are the root-mean-square radii of the neutron and proton 
density distributions in the nucleus, respectively. Neutron-rich nuclei having thick 
neutron skins are critically sensitive to the symmetry energy below nuclear 
matter saturation density~\cite{origin}.
If such a situation prevails in the outer crust of neutron stars, the predicted composition 
may be affected since the symmetry energy at subsaturation densities is less certain than at 
saturation density. Indeed, at the densities relevant for the description of the exotic (neutron-rich) 
nuclei forming the Coulomb lattice, the density dependence of the nuclear symmetry energy
is still not well constrained by the available experimental information.

\begin{figure}[t]
 \centerline{ \includegraphics[clip=true, width=0.6\linewidth]{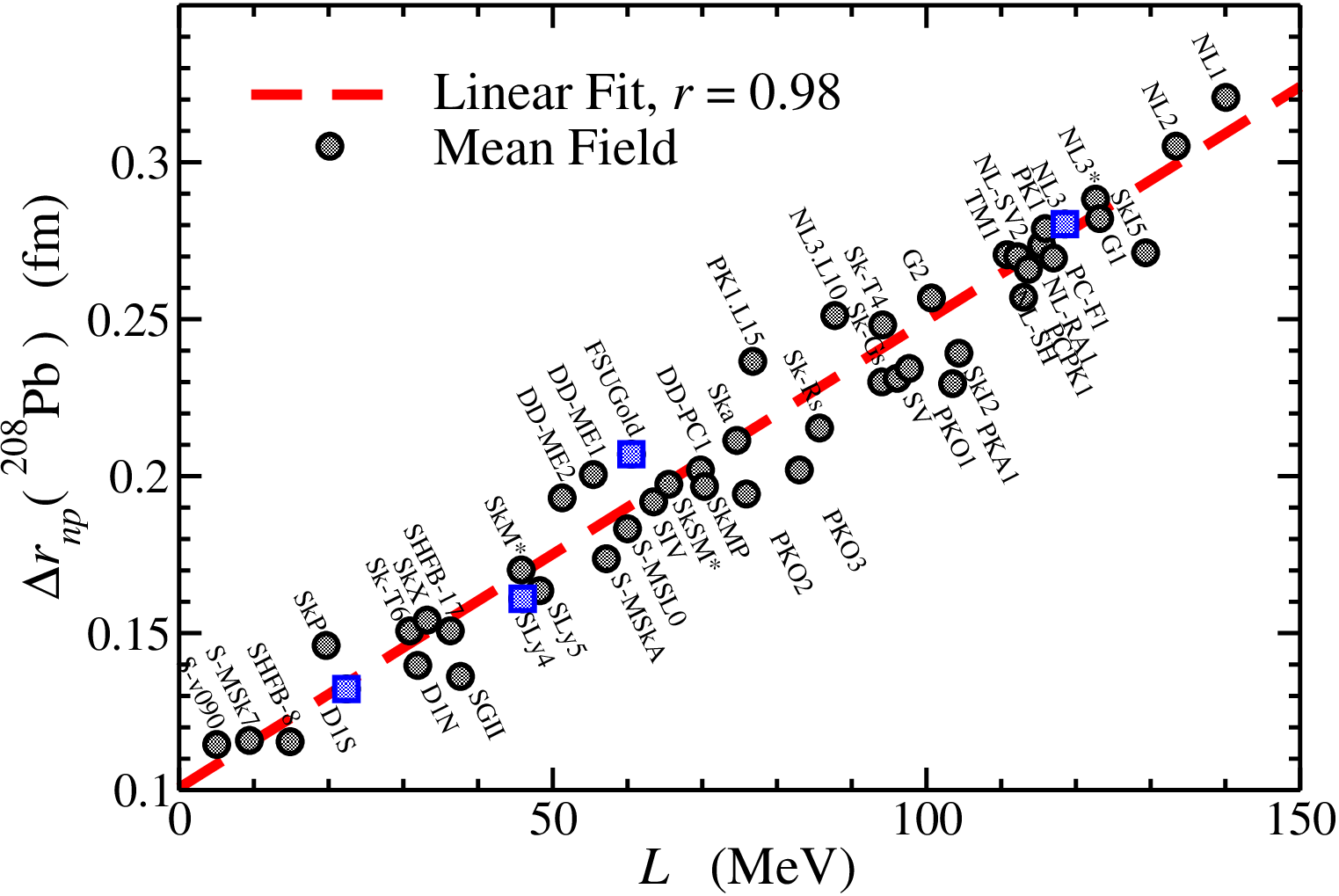}}
 \vspace{-0.2cm}
 \caption{The neutron skin thickness of ${}^{208}$Pb against the
           density slope of the nuclear symmetry energy at saturation
           as predicted by different nuclear mean-field interactions.
           The four interactions used in the calculations presented later in
           this chapter are shown by squares.}
 \label{Fig02}
\end{figure}

Our incomplete knowledge of the density dependence of the symmetry energy
also shows up in the microscopic effective nuclear mean-field (MF) models.
While they are accurate and reliable for the prediction of many ground-state
properties and excitation properties of finite nuclei along the whole periodic
table, the nuclear models usually differ in
the description of isospin-sensitive properties such as the neutron skin
thickness or the equation of state of asymmetric nuclear matter. It is
both practical and insightful to characterize the behavior of 
$E_{\rm sym}(\rho)$ in nuclear models by using a handful of parameters. 
To this end, it has become customary to expand $E_{\rm sym}(\rho)$
around the saturation density $\rho_0$ according to the formula
\begin{equation}
E_{\rm sym}(\rho) = J + L \Big( \frac{\rho-\rho_0}{3\rho_0} \Big) 
+ \frac{1}{2} K_{\rm sym} \Big( \frac{\rho-\rho_0}{3\rho_0} \Big)^2 
+ \cdots ,
\label{JLK}
\end{equation}
where $J$ is the value of the bulk symmetry energy and $L$ and 
$K_{\rm sym}$ describe, respectively, the slope and curvature of the 
symmetry energy at saturation. To emphasize the intimate connections
existing between the observables of neutron-rich nuclei and the
density-dependent symmetry energy in nuclear models, we show in
Fig.~\ref{Fig02} the correlation~\cite{Brown:2000, Furnstahl:2001un,
origin} between the neutron skin thickness $\Delta r_{np}$ of a
neutron-rich nucleus such as ${}^{208}$Pb and the parameter
\begin{equation}
L \equiv 3\rho_0\frac{\partial E_{\rm sym}(\rho)}{\partial\rho}
\Big|_{\rho_0} \ ,
 \label{l}
\end{equation}
which characterizes the slope of the symmetry energy at the saturation
density $\rho_0$. Note that the pressure of pure neutron matter at
saturation and the $L$ parameter are directly proportional in nuclear
models~\cite{KTAU}. The quantities $\Delta r_{np}$ of ${}^{208}$Pb and
$L$ displayed in Fig.~\ref{Fig02} have been calculated in a large and
representative set of nuclear MF models. These models are based on
very different theoretical grounds.
For example, in Fig.~\ref{Fig02} we have Skyrme zero-range forces (all
the models with names starting by S) and Gogny finite-range forces
(the D1S and D1N models). These are nonrelativistic forces derived
from an effective Hamilonian. Also displayed in Fig.~\ref{Fig02} are
relativistic forces based on effective field theory Lagrangians with
meson self-interactions, density-dependent meson-nucleon couplings,
and point couplings. The four interactions used in the calculations presented
later in this chapter are highlighted by squares.  
As it can be seen, these interactions cover the whole range of possible 
values for the $L$ parameter that is predicted by the large sample of MF 
calculations shown in Fig.~\ref{Fig02}. Irrespective of the nature of the models, 
all of them are able to describe similarly well general properties of nuclei
such as binding energies and charge radii along the valley of
stability. However, one clearly notes in Fig.~\ref{Fig02} a large
disagreement in the predictions for the size of the neutron skin
thickness of ${}^{208}$Pb, which ranges from about 0 fm to about 0.3
{\rm fm} in the various models. A large spread is also found in the
values of the $L$ parameter. Such discrepancies among models stem
mainly from the lack of accurate experimental information caused by
the inherent difficulties of probing uncharged, strongly interacting
particles (neutrons) model independently.

The novel ---and successfully commissioned--- Parity Radius Experiment
(PREx) at the Jefferson Laboratory~\cite{Kumar} has provided the first
model-independent evidence of the existence of a significant neutron
skin in ${}^{208}$Pb. PREx probes the neutron density distribution in
${}^{208}$Pb via parity-violating elastic electron
scattering~\cite{Kumar,Horowitz:1999fk}.  This experimental technique
is free of most strong interaction uncertainties,
similarly to elastic electron scattering for probing the proton density distribution in
nuclei. In this way, the experiment can constrain the neutron skin thickness of
$^{208}$Pb cleanly and, by means of the correlations shown in Fig.~\ref{Fig02}
and in Ref.~\cite{nostre}, the density dependence of the nuclear symmetry
energy. In recent years it has been demonstrated that a variety of neutron-star observables 
are correlated with the neutron skin thickness of ${}^{208}$Pb and the slope 
of the nuclear symmetry energy in nuclear models~\cite{Horowitz:2000xj,Horowitz:2001ya,Horowitz:2002mb, 
Carriere:2002bx,Steiner:2004fi,Steiner:2007rr}. One particularly interesting correlation of direct
relevance to the crustal region is a {\sl ``data-to-data''} relation between the neutron
skin thickness of ${}^{208}$Pb and the crust-to-core transition 
density~\cite{Horowitz:2000xj,COIMBRA}. 
Therefore, one may safely conclude that an improved knowledge of the isospin
properties of the nuclear interaction will impact deeply on the study of the crust of
neutron stars. Moreover, the benefits of these studies for the fields of nuclear structure and
nuclear astrophysics are mutual. For example, as pointed out at the beginning of this
Introduction, information deduced from observations of giant flares in strongly magnetized
neutron stars may prove to be an excellent opportunity for probing nuclei at extreme
conditions of density and isospin asymmetry that are hardly accessible
to terrestrial laboratories.

The present chapter has been organized as follows. The formalism required to compute the composition and the
equation of state of the outer crust is reviewed in Sec.~\ref{Formalism}. In Sec.~\ref{Results}
we employ several realistic nuclear-mass models to compute the structure and composition of the outer 
crust and focus on the sensitivity to the nuclear symmetry energy.
A more comprehensive study of the outer crust of nonaccreting cold neutron stars
can be found in the work by Ruester, Hempel, and
Schaffner-Bielich~\cite{Ruester:2005fm}. In spite of the fact that we do not use a large number of 
nuclear models for our investigations, we include a simple {\sl``toy model''} that provides critical 
insights into the role played by the symmetry energy. Moreover, in Sec.~\ref{Results} we
investigate the imprints of the density dependence of the symmetry energy on the sequence 
of neutron-rich nuclei occurring in the outer crust. Finally, our conclusions are 
laid in Sec.~\ref{Conclusions}

%
%
\section{Formalism}
\label{Formalism}
In the present section, we review the most important features of the formalism 
necessary to describe the outer crust of a nonaccreting neutron 
star~\cite{Roca-Maza:2008}. We closely follow the seminal ideas introduced by 
Baym, Pethick, and Sutherland~\cite{Baym:1971pw}. We 
refer the interested reader to Refs.~\cite{Haensel:1989,Haensel:1994,Ruester:2005fm}
for recent comprehensive studies.
 
The outer crust can be treated in good approximation as a cold system ($T\approx 0$)
that is composed by nuclei arranged in a Coulomb lattice and embedded
in a uniform free Fermi gas of electrons. The densities comprised by the outer crust 
go from the complete ionization of the electrons preserving electric neutrality
($\rho\!\approx\!10^{4} {\rm g/cm^{3}}$), to the neutron drip line ($\rho\!\approx\!10^{11}{\rm g/cm^{3}}$)
which defines the outer-inner crust interface (see Fig.~\ref{Fig01}). Therefore, the
composition of the outer crust is determined by the nucleus (with neutron number 
$N$, proton number $Z$, and baryon number $A\!=\!N\!+\!Z$) that minimizes ---for each
density--- the total energy per nucleon of the system. The energy per nucleon consists of 
the nuclear, electronic and lattice contributions:
\begin{equation}
 \varepsilon(A,Z;n)=\varepsilon_{n}+\varepsilon_{e}+
                    \varepsilon_{\ell}\;,
 \label{EoverA}
\end{equation}
where the baryon density is denoted by $n\!\equiv\!A_{\rm total}/V$. The 
nuclear contribution to the total energy per nucleon is
\begin{equation}
 \varepsilon_{n}(N,Z)\equiv\frac{M(N,Z)}{A}\;,  \;\;{\rm with}\;\;
  M(N,Z)=Nm_{n}+Zm_{p}-B(N,Z) \;.
 \label{EoverANucl}
\end{equation}
Note that within our description, the nuclear mass, $M(N,Z)$, does not 
depend on the density and, therefore, it will not contribute to the pressure 
of the system. $B(N,Z)$ is the binding energy of the nucleus and $m_{n}$ and $m_{p}$ are
neutron and proton masses, respectively.

The electronic contribution at the densities of interest is modeled
as a degenerate free Fermi gas~\cite{Baym:1971pw}. That is,
\begin{equation}
 \varepsilon_{e}(A,Z;n)=\frac{\mathscr{E}_{e}}{n}
  =\frac{1}{n\pi^{2}}\int_{0}^{p_{{\rm F}e}}
   p^{2}\sqrt{p^{2}+m_{e}^{2}}\,dp \;,
 \label{EoverAElec}
\end{equation}
where $\mathscr{E}_{e}$, $m_{e}$, and $p_{{\rm F}e}$ are the electronic 
energy density, mass, and Fermi momentum, respectively. Note that the 
electronic Fermi momentum and baryon density are related as follows:
\begin{equation}
 p_{{\rm F}e} = \left(3\pi^2 n_{e}\right)^{1/3}
             = \left(3\pi^2 y n\right)^{1/3} 
               \equiv y^{1/3}p_{\rm F}\;,
 \label{pFermie}
\end{equation}
where the electron fraction $y\!\equiv\!Z/A$ has been defined. 
In addition, for future convenience the following definition of 
the overall Fermi momentum has been introduced:
\begin{equation}
 p_{{\rm F}} = \left(3\pi^2 n\right)^{1/3} \;.
 \label{pFermi}
\end{equation}
The integral in Eq.~(\ref{EoverAElec}) can be evaluated
analytically and computed using the closed form
\begin{equation}
 \varepsilon_{e}(A,Z;n)=\frac{m_{e}^{4}}{8\pi^{2}n}
 \left[x_{\rm F}y_{\rm F}\Big(x_{\rm F}^{2}+y_{\rm F}^{2}\Big)
      -\ln(x_{\rm F}+y_{\rm F})\right]\;,
 \label{EoverAElec2}
\end{equation}
where a dimensionless Fermi momentum and energy have been defined as
follows:
\begin{equation}
  x_{\rm F}\!\equiv\!\frac{p_{{\rm F}e}}{m_{e}} 
        \;\;{\rm and}\;\;
  y_{\rm F}\!\equiv\!\frac{\epsilon_{{\rm F}e}}{m_{e}}=
          \sqrt{1+x_{\rm F}^{2}}\;.
 \label{peFermi}
\end{equation}

Finally, the last term in Eq.~(\ref{EoverA}) corresponds to the Coulomb lattice 
contribution to the total energy per particle. The completely ionized nuclei 
populating the outer crust feel the Coulomb repulsion among them and, at the
conditions present in this layer, crystallize in a body-centered-cubic lattice. Such a 
behavior has been demonstrated by Wigner for the case of an electron 
gas~\cite{Wigner:1934, Wigner:1938,Fetter:1971}. Accurate numerical calculations 
for the electron gas have been available for a long time~\cite{Coldwell:1960,Sholl:1967}
and the results can be readily generalized to the present 
case~\cite{Baym:1971pw}. We will take advantage of the previous investigations 
and use the expression for the lattice energy per nucleon as written in~\cite{Roca-Maza:2008}:
\begin{equation}
 \varepsilon_{\ell}(A,Z;n)=-
 C_{\ell} \frac{Z^{2}}{A^{4/3}}p^{}_{\rm F}
 \quad ({\rm with}\; C_{\ell}\!=\!3.40665\!\times\!10^{-3}) \;.
 \label{EoverALatt2}
\end{equation}
The full expression for the energy per baryon in terms of the 
nuclear, electronic, and lattice contributions is
\begin{equation}
 \varepsilon(A,Z;n)=\frac{M(N,Z)}{A}+
  \frac{m_{e}^{4}}{8\pi^{2}n}
  \left[x_{\rm F}y_{\rm F}\Big(x_{\rm F}^{2}+y_{\rm F}^{2}\Big)
       -\ln(x_{\rm F}+y_{\rm F})\right]
       -C_{\ell} \frac{Z^{2}}{A^{4/3}}p^{}_{\rm F} \;.
 \label{EoverA2}
\end{equation}
Note that all but the nuclear contribution, $M(N,Z)$, to the total energy 
per baryon are well known by the current theory. Such an 
uncertainty is due to the fact that it is very difficult to deal with a many-body 
system of strongly interacting fermions even though the underlying theory,
Quantum Chromodynamics, was established many years ago. This affects
the calculations of the structure and composition of 
the outer crust since one cannot avoid the use of nuclear mass models. Experimental 
data on nuclear masses are available for a large number of nuclei around the line
of stability but, unfortunately, masses of nuclei with large isospin asymmetries such
as some of the nuclei that may exist in the outer crust are unknown. Therefore,
observational information sensitive to crustal properties of neutron stars can provide
valuable insights into exotic nuclei unexplored in terrestrial laboratories.
Reciprocally,
the advent of new facilities capable to produce rare ion beams for experimental
studies in the laboratory may place strong constraints on the crustal properties.

In modeling the outer crust, the central assumption is that of thermal, hydrostatic, 
and chemical equilibrium. For this reason, we also calculate the equation of state
(namely, the relation between pressure and density) and the chemical
potential, since complete equilibrium demands the equality of temperature,
pressure, and chemical potential at each layer of the outer crust. As we have mentioned, 
the individual nuclei do not contribute to the pressure 
and, then, only the electronic and lattice terms contribute:
\begin{eqnarray}
  P(A,Z;n)&=& -\left(\frac{\partial E}{\partial V}\right)_{\!\!A,Z} 
          \nonumber\\
          &=&\frac{m_{e}^{4}}{3\pi^{2}}\left(x_{\rm F}^{3}y_{\rm F} 
            - \frac{3}{8}\left[x_{\rm F}y_{\rm F}\Big(x_{\rm F}^{2}+
              y_{\rm F}^{2}\Big)-\ln(x_{\rm F}+y_{\rm F})\right]\right)
            -\frac{n}{3}C_{\ell} \frac{Z^{2}}{A^{4/3}}p^{}_{\rm F}\;.
          \nonumber\\
 \label{PTotal}
\end{eqnarray}

The temperature of the system is in good approximation assumed to be equal to zero 
(on the scale of nuclear energies) and, therefore, the only remaining thermodynamic 
observable to calculate is the chemical potential. At zero temperature, the Gibbs
free energy and the total energy of the system are related by a
Legendre transform ($G\!=\!E\!+\!PV$). That is,
\begin{equation}
 \mu(A,Z;P)=\frac{G(A,Z;P)}{A_{\rm total}}  
           =\varepsilon(A,Z;n)+\frac{P}{n}
           =\frac{M(N,Z)}{A}+\frac{Z}{A}\mu_{e}
           -\frac{4}{3}C_{\ell}\frac{Z^{2}}{A^{4/3}}p^{}_{\rm F}\;,
 \label{ChemPotential}
\end{equation}
where $\mu_{e}\!=\!\sqrt{p_{{\rm F}e}^{2}+m_{e}^{2}}$ is the
electronic chemical potential. Note that the chemical potential is a
function of the pressure whereas the energy per baryon is a function
of the baryon density. The transformation from one into the other is
accomplished through Eq.~(\ref{PTotal}). Actually, it is convenient to compute 
the composition of the outer crust by minimizing the chemical potential 
at a constant pressure rather than by minimizing the energy per particle 
at constant baryon density. This is because the equilibrium conditions 
demands the pressure and chemical potential be continuous throughout 
the outer crust but this do not necessarily imply the same for the baryon 
density and energy.

\section{Results}
\label{Results}
Following the formalism presented in the previous section, we provide here 
the results obtained for the structure and composition of the outer crust 
of a nonaccreting neutron star. First, we develop a ``toy-model'' with the aim 
of understanding the role of the different terms in the energy and 
chemical potential. Secondly, we use for the calculations two of the most accurate 
mass models available in the literature. For this reason, 
we will take them as a reference along this section. These two models are
the one by Duflo and Zuker~\cite{Duflo:1994,Zuker:1994, Duflo:1995} and the
finite range droplet model of M\"oller, Nix and 
collaborators~\cite{Moller:1993ed,Moller:1997bz}. 
Both models are based on sophisticated 
microscopic/macroscopic approaches that yield root-mean-square
(rms) errors of only a fraction of an {\rm MeV} when
compared to large databases of available experimental nuclear 
masses~\cite{Audi:1993zb,Audi:1995dz}. 
First and foremost, we are interested in understanding how successful models
in describing ground-state properties of stable nuclei differ in their predictions of
exotic (neutron-rich) nuclei and in correlating such differences to the little constrained
isovector channel of those models. Unfortunately, the microscopic/macroscopic
mass models do not provide predictions for the nuclear symmetry energy at different
densities. This fact motivates the use of nuclear MF models that are also accurate in 
the description of nuclear masses ---typical rms deviations are of a few {\rm MeV} when compared to
large databases of known masses--- and predict a specific density dependence of the nuclear
symmetry energy. Our set of MF models has been selected to represent a broad range of values for
the $L$ parameter (see Fig.~\ref{Fig02}) that describes the density slope of the nuclear symmetry
energy at saturation.

\subsection{A Toy Model of the Outer Crust}
\label{ToyModel}

We first introduce a simple 
{\sl ``toy model''} that captures the essential physics 
of the outer crust. That is, the competition between an electronic density 
that drives the system towards more neutron-rich nuclei and a nuclear 
symmetry energy that opposes such a change.
The toy model is based on the following two approximations. First, a
simple liquid-drop model will be used to compute nuclear masses [see
Eq.~(\ref{EoverANucl})]. Second, the electronic contribution will be
assumed to be that of an extremely relativistic ({\it i.e.},
$m_{e}/p_{{\rm F}e}\!\rightarrow\!0$) Fermi gas. Both of these
approximations will allow an analytic treatement and provide a not too 
simplified approach to the outer crust. For these reasons, 
valuable information on the physics taking place in the outer crust will be deduced.

In the absence of pairing correlations, the liquid-drop mass formula may be written
as follows:
\begin{equation}
 \varepsilon_{n}(x,y) = 
  m_{p}y + m_{n}(1-y) - a_{\rm v} + \frac{a_{\rm s}}{x} +
  a_{\rm c}x^{2}y^{2} + a_{\rm a}(1-2y)^{2} \;,
 \label{LiquidDrop}
\end{equation}
where $x\!\equiv\!A^{1/3}$ and $y\!\equiv\!Z/A$. The various empirical 
constants ($a_{\rm v}$, $a_{\rm s}$, $a_{\rm c}$, and $a_{\rm a}$) represent 
volume, surface, Coulomb, and asymmetry contribution, respectively. Using a 
least-squares fit to 2049 nuclei 
(available online at the UNEDF collaboration website {\tt http://www.unedf.org/}) 
one obtains the following values for the four empirical constants:
\begin{equation}
  a_{\rm v}=15.71511~{\rm MeV}, \;
  a_{\rm s}=17.53638~{\rm MeV}, \;
  a_{\rm c}= 0.71363~{\rm MeV}, \;
  a_{\rm a}=23.37837~{\rm MeV}.
 \label{LiquidDropParams}
\end{equation}

At zero density, the only term contributing to the total energy of the 
outer crust is $M(N,Z)$. In this situation, the optimal values of
$x$ and $y$ using the simple liquid-drop formula can be calculated by setting 
both derivatives in Eq.~(\ref{LiquidDrop}) equal to zero. The simple analytic
solution is
\begin{subequations}
\begin{align} 
   &   A=x^{3}=\left(\frac{a_{\rm s}}{2a_{\rm c}}\right)\frac{1}{y^{2}}\;, 
     \label{dEpsilonxSol}\\ 
   &   y=\frac{1+\displaystyle{\left(\frac{\Delta m}{4a_{\rm a}}\right)}}
       {2+\displaystyle{\left(\frac{a_{\rm c}}{2a_{\rm a}}\right)x^{2}}} 
       \approx \frac{1/2}{1+\displaystyle{\left(\frac{a_{\rm c}}
       {4a_{\rm a}}\right)x^{2}}}\;,
     \label{dEpsilonySol}
 \end{align} 
 \label{dEpsilonSol}
\end{subequations}
where we have defined $\Delta m\!\equiv\!m_{n}\!-\!m_{p}$. 
The above solutions suggest that for a fixed proton fraction 
$y\!=\!Z/A$, the optimal value of $x$
emerges from a competition between surface (which favors large $x$)
and Coulomb contributions (which favors small $x$). 
Conversely, if $A\!=\!x^{3}$ is held fixed, then the optimal proton
fraction $y$ results from the competition between Coulomb and
asymmetry contribution, with the former favoring $y\!=\!0$ and
the latter $y\!=\!1/2$. When both equations are solved 
{\sl simultaneously}, one finds the most stable nucleus for
this parameter set: $x_{0}\!=\!3.906$ and 
$y_{0}\!=\!0.454$, or equivalently: $A_{0}=59.598$, $Z_{0}=27.060$, 
$N_{0}=32.538$ and $(B/A)_{0}=8.784~{\rm MeV}$.

The second assumption defining the toy model is that of an
utrarelativistic Fermi gas of electrons ({\it i.e.}, $p_{{\rm F}e}\!\gg\!m_{e}$).
In this limit one obtains simple expressions for the total energy per 
baryon, chemical potential, and pressure in terms of the adopted set of 
variables. That is,
\begin{subequations}
\begin{align} 
  &   \varepsilon(x,y,p_{\rm F}^{}) = \varepsilon_{n}(x,y) +
    \frac{3}{4}y^{4/3}p_{\rm F}^{}-C_{\ell}\,x^{2}y^{2}p^{}_{\rm F}\;,
  \label{Energy} \\
  &   \mu(x,y,p_{\rm F}^{}) = \varepsilon_{n}(x,y) + y^{4/3}p^{}_{\rm F}
           -\frac{4}{3}C_{\ell}\,x^{2}y^{2}p^{}_{\rm F}\;,
    \label{ChemPot}\\
  &   P(x,y,p_{\rm F}^{}) = \frac{n}{4}y^{4/3}p^{}_{\rm F}
  -\frac{n}{3}C_{\ell}\,x^{2}y^{2}p^{}_{\rm F}\;.
    \label{Pressure}
  \end{align}
 \label{EPandMu}
\end{subequations}
Assuming a neutron-drip density of about $4\!\times\!10^{11}{\rm
g/cm}^{3}$ consistently found in all calculations (see Table~\ref{Table1} below),
the overall Fermi 
momentum is approximately equal to $p_{\rm F}^{\rm drip}\!\approx\!40$~{\rm MeV}. 
This provides a large electronic contribution at the outer-inner crust 
interface of about $\varepsilon_{e}^{\rm drip}\!\approx\!30\,y^{4/3}_{\rm drip}\approx 6.2$~{\rm MeV}
when one takes the proton fraction of the conventionally accepted drip nucleus ${}^{118}$Kr.
Therefore, the electrons will drive the system to small
values of $y$ as compared to the solution at zero density ($y_{0}\!=\!0.454$)
since the lattice contribution has a minor effect on the total energy:
$\varepsilon_{l}^{\rm drip}\!\approx\!-0.14\,y^2_{\rm drip}x^2_{\rm drip}\approx -0.3$~{\rm MeV}. 
Unsurprisingly, the lattice contribution has the same dependence in the parameters $x$
and $y$ as the Coulomb term in the liquid-drop mass formula. Hence, it can 
be included through a density dependent redefinition 
of the Coulomb coefficient: $a_{\rm c}\!\rightarrow\!\widetilde{a}_{c}
(p^{}_{\rm F}) \!\equiv\! a_{\rm c}-C_{\ell}p^{}_{\rm F}$.

The analytical equations to be solved when the electronic and lattice
contributions are incorporated to the liquid-drop mass formula in
order to describe the total energy of the system are, at a fixed
density and in terms of the $x$ and $y$ variables,
\begin{subequations}
\begin{align} 
   &   \left(\frac{\partial\varepsilon}{\partial x}
           \right)_{\!\!\!y,p^{}_{\rm F}} = 
          -\frac{a_{\rm s}}{x^{2}}+
            2\widetilde{a}_{\rm c}xy^{2}=0\;,
           \label{dEpsilonx2}\\ 
   &   \left(\frac{\partial\varepsilon_{n}}{\partial y}
           \right)_{\!\!\!x,p^{}_{\rm F}} = 
          -\Delta m+2\widetilde{a}_{\rm c}x^{2}y - 4a_{\rm a}(1-2y)
          +y^{1/3}p_{\rm F}^{}=0\;.
           \label{dEpsilony2}
  \end{align}
 \label{dEpsilon2}
\end{subequations}
Therefore, the optimal value of the proton fraction $y$ (for fixed $x$) will emerge 
from a competition between the redefined Coulomb and asymmetry terms where the 
lattice contribution corrects the solution at zero density towards a more
symmetric $y\!=\!1/2$ configuration. The electron gas is responsible 
for driving the system towards progressively more neutron-rich nuclei. Thus, the
outer crust represents a unique laboratory for the study of neutron-rich nuclei in
the $Z\!\approx\!20\!-\!50$ region. With the advent of new 
{\sl Rare-Isotope Facilities} worldwide that aim to provide a detailed
map of the nuclear landscape, observational and theoretical studies of the neutron 
star crust can constitute a very important complement.

\subsubsection{First-order Solution}
\label{FirstOrder}

The exact solutions of Eqs.~(\ref{dEpsilon2}) are, unfortunately, not 
analytical. For that reason, we first compute
approximate solutions accurate to first order in $p_{\rm F}^{}$.  
The analytic first-order solutions have the advantage of
providing valuable insights into the composition of the outer crust. 
Those solutions are
\begin{subequations}
 \begin{align}
 & x(p^{}_{\rm F})=x_{0}\left[1+
  \left(\frac{(C_{1}-1)C_{\ell}+2C_{2}}{3C_{1}-1}\right)
  \frac{p_{\rm F}}{a_{\rm c}}\right] =
  (3.90610+0.03023p^{}_{\rm F}) \;,
  \label{FirstOrderSolsx}\\ 
 & y(p^{}_{\rm F})=y_{0}\left[1-
  \left(\frac{3C_{2}-C_{\ell}}{3C_{1}-1}\right)
   \frac{p_{\rm F}}{a_{\rm c}}\right] =
  (0.45405-0.00419p^{}_{\rm F})\;.
  \label{FirstOrderSolsy}
 \end{align}
 \label{FirstOrderSols}
\end{subequations}
Note that in the above expressions the Fermi momentum is 
given in MeV. Moreover, for simplicity, the following two 
dimensionless quantities were introduced:
\begin{equation}
  C_{1} \equiv \frac{4a_{a}}{x_{0}^{2}a_{\rm c}} \approx 8.58843
  \;\;\; {\rm and} \;\;\;
  C_{2} \equiv \frac{1}{2x_{0}^{2}y_{0}^{2/3}} \approx 0.05547 \;.
\end{equation}

What was previously discussed is now qualitatively confirmed by 
the first-order solutions of the ``toy-model''. That is, the 
proton fraction $y$ decreases with density in an effort to minimize 
the ``repulsive'' electronic contribution. An excellent approximation 
of Eq.~(\ref{FirstOrderSolsy}) is 
\begin{eqnarray}
 y(p^{}_{\rm F})&=&y_{0}-\frac{p^{}_{{\rm F}e}}{8a_{\rm a}}\nonumber\\
              &\approx&y_{0}- y_{0}^{1/3} \frac{p_{\rm F}}{8a_{\rm a}}
              \;= \; 0.45405-0.00411p^{}_{\rm F} \;.  
 \label{FirstOrderSolsy2}
\end{eqnarray}
In the latter expression, the denominator ($8a_{\rm a}$)
is significantly larger than the electronic 
Fermi momentum over the entire region of interest. Specifically,  
$p_{{\rm F}e}/8a_{\rm a} \leq 0.00411 p^{drip}_{{\rm F}}\approx 1 / 6$. 
For this reason, we expect that the first-order approximation is fairly accurate 
over the entire outer crust. Indeed, assuming the last approximated expression and a drip 
density of $\rho_{\rm drip}\!=\!4\!\times\!10^{11}~{\rm g/cm}^{3}$ one finds a
proton fraction of $y^{}_{\rm drip}\!=\!0.298$ which represents a 2\%
deviation from the value of $y({}^{118}{\rm Kr})\!=\!0.305$ ---for the
conventionally accepted drip nucleus ${}^{118}$Kr. Thus, if Eq.~(\ref{FirstOrderSolsy2})
is confirmed to be accurate when compared to the exact solution of the ``toy-model'' 
and to the more realistic calculations of the following section, it will provide a very simple 
and clear picture for the evolution of the proton fraction througout the outer crust. That is, the larger 
the value of the asymmetry energy coefficient $a_{\rm a}$, the slower the evolution away from 
$y_{0}$ and, therefore, the more symmetric the nuclei in the crustal lattice will remain.

\begin{figure}[h!]
  \centerline{ \includegraphics[width=0.6\linewidth]{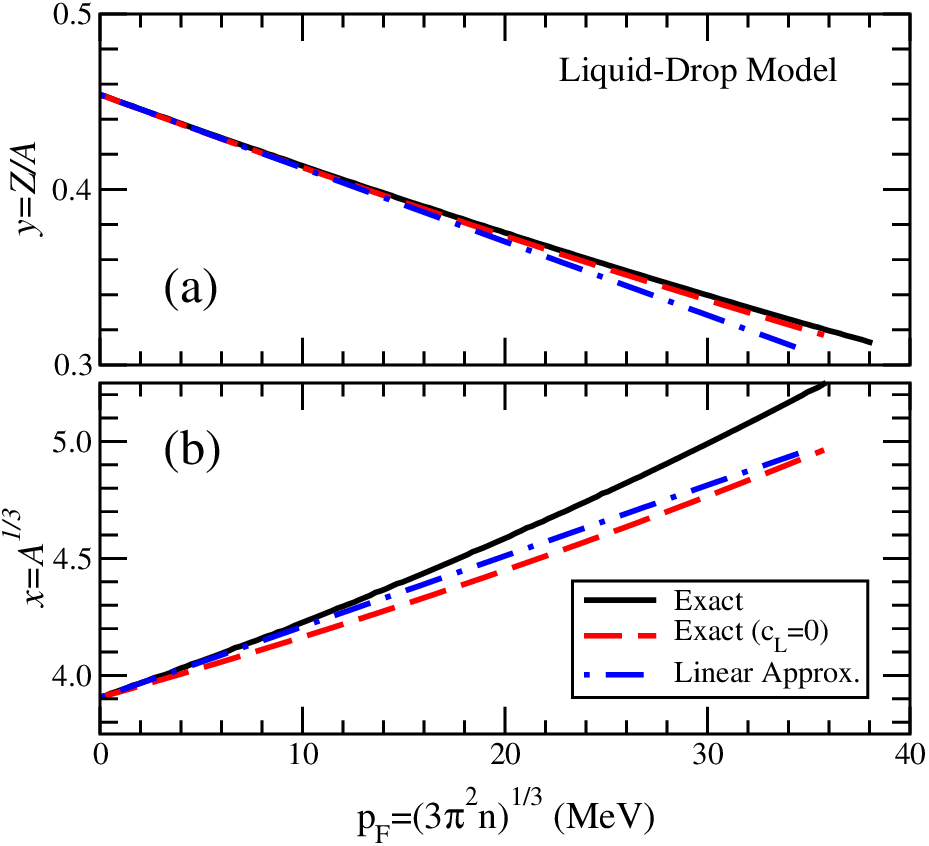}}
 \vspace{-0.2cm}
 \caption{
          Proton fraction $y\!=\!Z/A$ (a) and baryon number 
          $x\!=\!A^{1/3}$ (b) are displayed as a function of 
          the Fermi momentum 
          $p^{}_{\rm F}\!\equiv\!(3\pi^{2}n)^{1/3}$. The black 
          solid lines represent the exact solution to the 
          toy-model problem given in Eqs.~(\ref{dEpsilon2}), 
          while the red dashed lines display the corresponding
          solution in the $C_{\ell}\!\equiv\!0$ (no lattice) 
          limit [see Eqs.~(\ref{ToySols})]. Finally, the 
          low-density solution [Eqs.~(\ref{FirstOrderSols})] 
          is displayed by the blue dot-dashed lines.} 
 \label{Fig03}
\end{figure}

\subsubsection{Exact Solution}
\label{ExactToySolution}

Although numerically simple, the exact solution of the toy-model 
problem for $x$ and $y$ at a fixed density cannot be displayed in 
an analytical form. However, there is the possibility of presenting 
an approximate closed solution slightly different from the exact one 
in terms of $x$ and the overall Fermi momentum $p_{\rm F}$ when the 
lattice contribution is neglected. That is,
\begin{subequations}
\begin{align}
   &   x(y)=\left(\frac{a_{\rm s}}{2{a_{\rm c}}y^{2}}\right)^{1/3}\;,
           \label{ToySols1} \\ 
   &   p^{}_{\rm F}(y)=\frac{\Delta m-2a_{\rm c}x^{2}y+4a_{\rm a}(1-2y)}
                    {y^{1/3}} \;.  
           \label{ToySols2}
 \end{align}
 \label{ToySols}
\end{subequations}
These equations suggest that one can solve analyticaly the problem
for $x$ and $p^{}_{\rm F}$ as a function of $y$, with the maximum 
value of $y$ given by $y_{\rm max}\!=\!y_{0}\!=\!0.45405$ and the minimum 
value of $y$ given by the neutron drip-line condition 
$\mu(y_{\rm min})\!=\!m_{n}$. 

In Fig.~\ref{Fig03} the baryon number $x\!=\!A^{1/3}$ and proton
fraction $y\!=\!Z/A$ are displayed as a function of the Fermi 
momentum $p^{}_{\rm F}\!\equiv\!(3\pi^{2}n)^{1/3}$. The black solid 
lines display the exact numerical solution to the toy-model problem
[see Eqs.~(\ref{dEpsilon2})]. In this simple model, the drip line 
density is predicted to be at 
$\rho_{\rm drip}\!=\!4\!\times\!10^{11}~{\rm g/cm}^{3}$ with the 
drip-line nucleus being ${}^{154}$Cd ({\it i.e.}, $Z\!=\!48$ and 
$N\!=\!106$). The solution obtained by ignoring the lattice
contribution is displayed by the red dashed lines. Because the
lattice contribution to the chemical potential is negative, the
$C_{\ell}\!\equiv\!0$ solution reaches the drip line faster,
{\it i.e.}, at a lower density. Moreover, as the lattice
contribution {\sl ``renormalizes''} the Coulomb term in the
semi-empirical mass formula (or equivalently, enhances the role 
of the symmetry energy) the $C_{\ell}\!\equiv\!0$ solution predicts 
a lower proton fraction than the exact solution. Finally, the 
dot-dashed blue lines display the solution correct to first-order
in $p^{}_{\rm F}$. In the particular case of the proton fraction 
$y$, the approximate linear solution 
$y=y_{0}-p^{}_{{\rm F}e}/8a_{\rm a}$ [Eq.~(\ref{FirstOrderSolsy2})]
reproduces fairly accurately the behavior of the exact solution.

\begin{figure}[t]
  \centerline{ \includegraphics[width=0.6\linewidth]{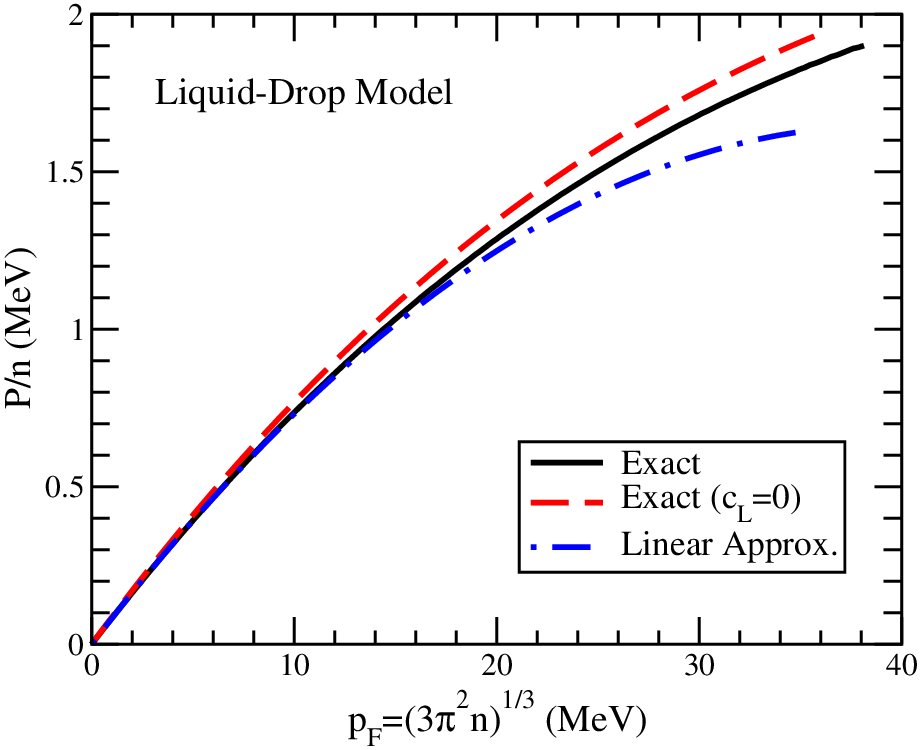}}
 \vspace{-0.2cm}
 \caption{
          Pressure as a function of the Fermi momentum 
          $p^{}_{\rm F}\!\equiv\!(3\pi^{2}n)^{1/3}$. The black 
          solid line represents the exact solution to the
          toy-model problem, the red dashed line displays 
          the corresponding $C_{\ell}\!\equiv\!0$ (no lattice)
          solution, and the low-density solution is displayed 
          by the blue dot-dashed line.}
 \label{Fig04}
\end{figure}

The equation of state ({\it i.e.}, pressure {\it vs} density)  
predicted by the toy model is displayed in Fig.~\ref{Fig04}. 
As the lattice provides a negative contribution to the pressure 
[Eq.~(\ref{Pressure})], the equation of state for the 
$C_{\ell}\!\equiv\!0$ case is slightly stiffer than the 
exact one. The first-order solution in $p^{}_{\rm F}$ provides 
a {\sl quantitatively accurate} description of the equation of 
state up to fairly large values of the density. Note that the 
first-order approximation to the pressure is defined as follows:
\begin{equation}
 \frac{P}{np^{}_{\rm F}}=\frac{1}{4}y^{4/3}
           -\frac{1}{3}C_{\ell}\,x^{2}y^{2}
 \approx (0.08367-0.00106p^{}_{\rm F})
\;.  
 \label{FirstOrderP}
\end{equation}

\subsection{Realistic Models of the Outer Crust}
\label{RealisticModel}
 
In this section we employ realistic nuclear mass models to compute the
structure and composition of the outer crust. Two of the
models~\cite{Moller:1993ed,Moller:1997bz,Duflo:1994,Zuker:1994,Duflo:1995}
are based on sophisticated mass formulas that have been calibrated to
the around two thousand available experimental masses throughout the periodic
table~\cite{Audi:1993zb,Audi:1995dz}. The other four models are based
on accurately-calibrated microscopic approaches that employ a handful 
of empirical parameters to reproduce the ground-state properties of 
finite nuclei and some nuclear collective 
excitations~\cite{Lalazissis:1996rd,Lalazissis:1999,
Todd-Rutel:2005fa,D1S,SLY4,SLY4b}.

The microscopic MF models, although not as accurate as the
microscopic/macroscopic ones in the description of nuclear masses, provide
useful insights on various details of the underlying physics. The MF models
typically have around 10 parameters adjusted to reproduce some selected
nuclear data. In contrast to the microscopic/macroscopic models, which
have usually a much larger number of parameters, the microscopic
models are solved self-consistenly and, therefore, the bulk and
surface contributions to the total energy are closely related. This
situation allows one to unravel possible correlations between the
predicted finite nuclei properties and the properties of the infinite
system, such as the correlation between the neutron skin thickness of
lead and the slope of the symmetry energy shown in Fig.~\ref{Fig02}.
In this sense, the critical role played by the symmetry energy in the
evolution of the proton fraction with density can be studied within
the framework of the mean-field approach.

\begin{figure}[t]
  \centerline{ \includegraphics[width=0.6\linewidth]{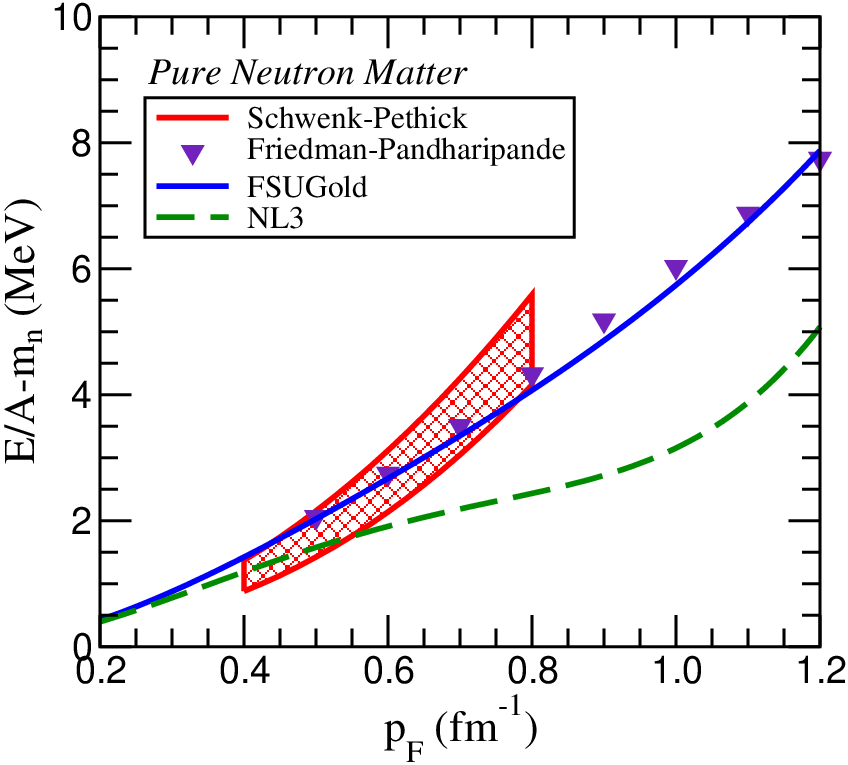}}
 \vspace{-0.2cm}
 \caption{ Energy per particle for pure neutron matter
         as a function of the neutron Fermi momentum. Shown are 
         the microscopic model of Friedman and Pandharipande
         ~\cite{Friedman:1981qw} (purple triangles) and the 
         {\sl model-independent} result based on the physics 
         of resonant Fermi gases by Schwenk and Pethick
         ~\cite{Schwenk:2005ka} (red region). Also shown are
         the predictions from the accurately calibrated
         NL3~\cite{Lalazissis:1996rd,Lalazissis:1999} (green 
         dashed line) and  FSUGold~\cite{Todd-Rutel:2005fa} 
         (blue line) models.}
 \label{Fig05}
\end{figure}

\begin{figure}[t]
  \centerline{ \includegraphics[clip=true, width=0.6\linewidth]{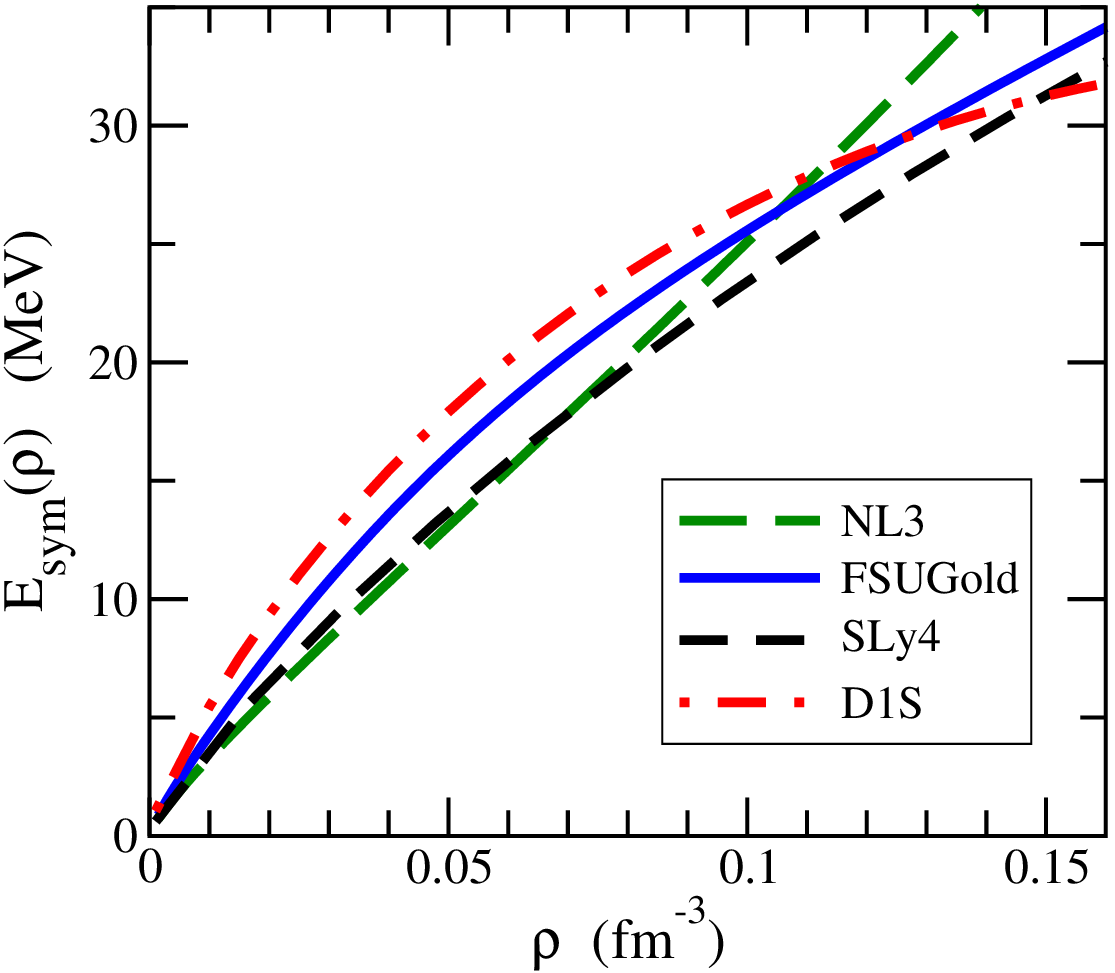}}
 \vspace{-0.2cm}
 \caption{Density dependence of the symmetry energy predicted
          at nuclear subsaturation by the indicated microscopic
          mean-field models.}
 \label{Fig06}
\end{figure}

We do not know from experiment how the symmetry energy coefficient
$a_a$ changes as nuclei approach the neutron drip line. Nuclear mean
field models predict the development of a significant neutron skin
that renders these nuclei more diffuse. If so, one needs to extrapolate
the symmetry energy to lower densities, a procedure that is highly
uncertain because of our poor knowledge of the density derivative of
the symmetry energy.
To illustrate this uncertainty, the equation of state of pure neutron matter predicted 
by NL3 (green dashed line) and FSUGold (blue solid line) is displayed 
in Fig.~\ref{Fig05}.  For comparison, we also show the predictions
from the microscopic model of Friedman and Pandharipande based on 
realistic two-body interactions~\cite{Friedman:1981qw} (purple 
upside-down triangles) and the \emph{model-independent} result 
based on the physics of resonant Fermi gases by Schwenk and 
Pethick~\cite{Schwenk:2005ka} (red hatched region). Note that to a
very good approximation, the equation of state of pure neutron matter
equals that of symmetric nuclear matter \emph{plus} the symmetry
energy~\cite{KTAU}. The differences between NL3 and FSUGold displayed in
Fig.~\ref{Fig05} are {\sl all} due to the large uncertainties
in the symmetry energy. In particular, as NL3 predicts a stiffer 
equation of state than FSUGold, namely, one whose energy increases 
faster with density {\sl at suprasaturation densities}, the symmetry
energy of NL3 is lower than that of FSUGold {\sl at subsaturation
densities} (see Fig.~\ref{Fig06}). Thus, FSUGold has been shown to
reach the neutron-drip lines earlier than NL3~\cite{Todd:2003xs}. By
the same token, NL3 should predict a sequence of more neutron-rich
nuclei (lower $y$) in the outer crust than FSUGold.

We depict in Fig.~\ref{Fig06} the behavior of the symmetry energy at
subnormal nuclear densities as predicted by the microscopic mean-field
models, including the nonrelativistic interactions SLy4 and D1S in
addition to the covariant models NL3 and FSUGold. The value of the
bulk symmetry energy at saturation density is $J= 37.4, 32.6, 32.0$,
and 31.2~MeV for NL3, FSUGold, SLy4, and D1S, respectively, whereas
the value of the slope of the symmetry energy at saturation [recall
Eqs.~(\ref{JLK}) and (\ref{l})] is $L= 118, 60, 46$, and 22~MeV for
the same four models. We observe that in spite of the fact that NL3
has the largest symmetry energy at saturation ($J= 37.4$ MeV), it
predicts the lowest symmetry energy at subsaturation densities
$\rho\lesssim 0.1$~fm$^{-3}$, owing to the fact that NL3 has (by far)
the largest $L$ value in the four models. Indeed, in the region
$\rho\lesssim 0.1$~fm$^{-3}$ the relativistic NL3 parameter set and
the SLy4 Skyrme interaction yield a quite similar symmetry energy. The
D1S Gogny force, having the lower $J$ and $L$ values in the considered
parameter sets, predicts the largest symmetry energy in the regime
$\rho\lesssim 0.1$~fm$^{-3}$. The results by the relativistic
parameter set FSUGold in the alluded density region are not far from
the D1S curve. Thus, we advance that one may expect more similar
predictions for the composition of the outer crust between NL3 and
SLy4, on the one hand, and between FSUGold and D1S, on the other hand.

\begin{figure}[t]
  \centerline{ \includegraphics[width=0.6\linewidth]{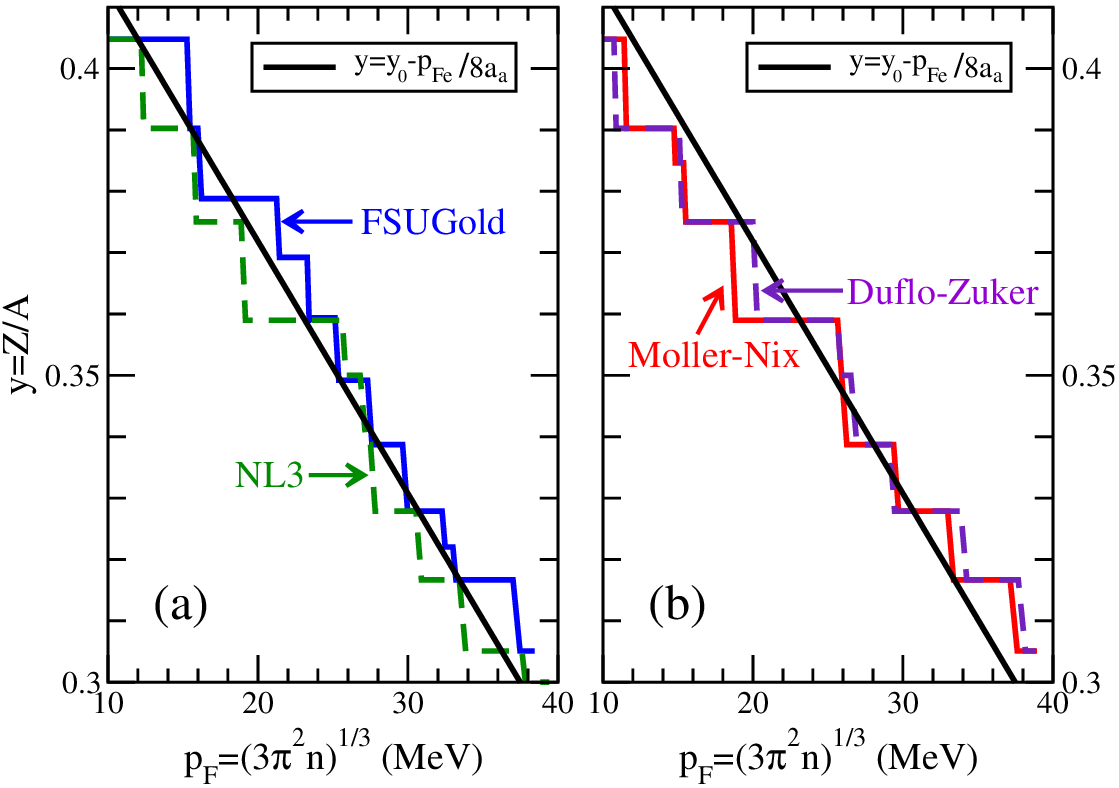}}
 \vspace{-0.2cm}
 \caption{ Panel (a) displays the
          proton fraction predicted by the accurately calibrated 
          FSUGold (blue solid line) and NL3 (green dashed line) 
          parameter sets. Also shown is the simple liquid-drop 
          formula given in Eq.~(\ref{FirstOrderSolsy2}).
          Panel (b) shows the proton fraction predicted by the
          Moller-Nix (red solid line) and Duflo-Zuker (purple dashed line)
          mass formulas.}
 \label{Fig07}
\end{figure}

Shown in the left-hand panel of Fig.~\ref{Fig07} is the proton fraction  
predicted by the two microscopic models FSUGold (blue solid line) and
NL3 (green dashed line). Also shown is the estimate
obtained from the liquid-drop formula [Eq.~(\ref{FirstOrderSolsy2})].
In spite of its simplicity, Eq.~(\ref{FirstOrderSolsy2}) nicely
averages the result of the realistic calculations, which confirms its
usefulness for a general and qualitative understanding of the
composition of the outer crust of a neutron star.
The proton fraction predicted with the FSUGold parameter set is
consistently higher than for the NL3 set. This is a reflection of the
stiffer penalty imposed on the FSUGold set for departing from the
symmetric ($N\!=\!Z$) limit. The right-hand panel shows the 
corresponding behavior for the case of the microscopic/macroscopic
models of Moller-Nix (red solid line) and Duflo-Zuker (purple dashed
line). Differences among these models are small.

\begin{figure}[t]
  \centerline{ \includegraphics[width=0.6\linewidth]{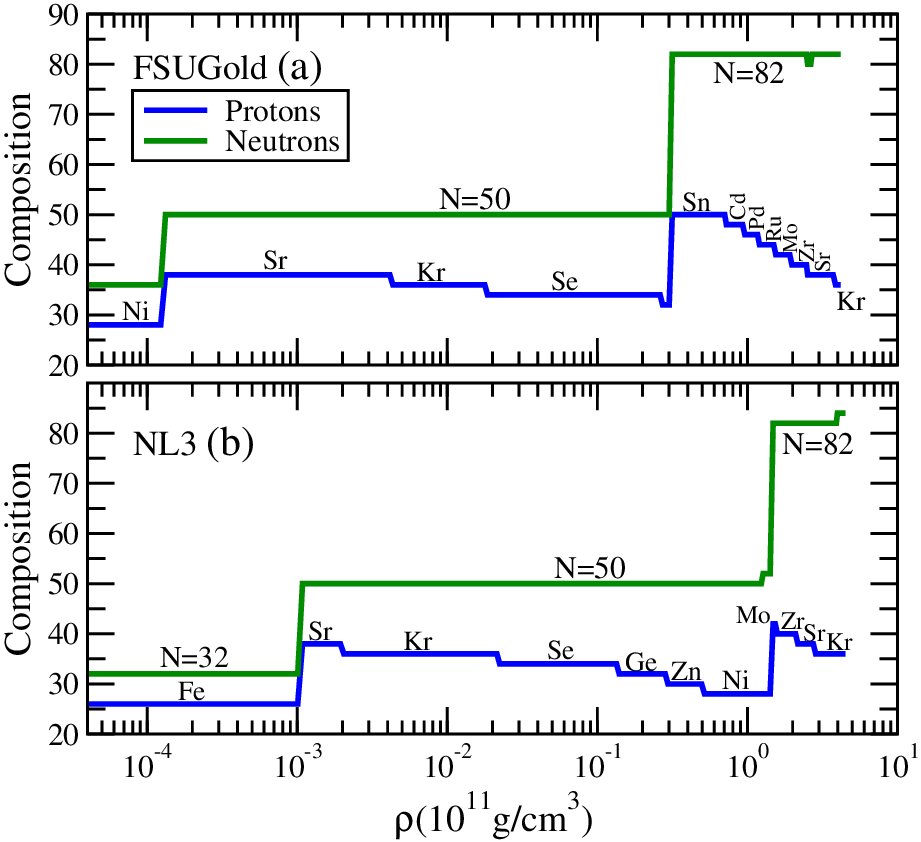}}
 \vspace{-0.2cm}
 \caption{ Composition of the outer crust of a 
           neutron star as predicted by the relativistic mean-field
           parameter sets FSUGold (a) and NL3 (b). Protons are
           displayed with the (lower) blue line while neutrons 
           with the (upper) green line.}
 \label{Fig08}
\end{figure}

\begin{figure}[h!]
  \centerline{ \includegraphics[width=0.6\linewidth]
{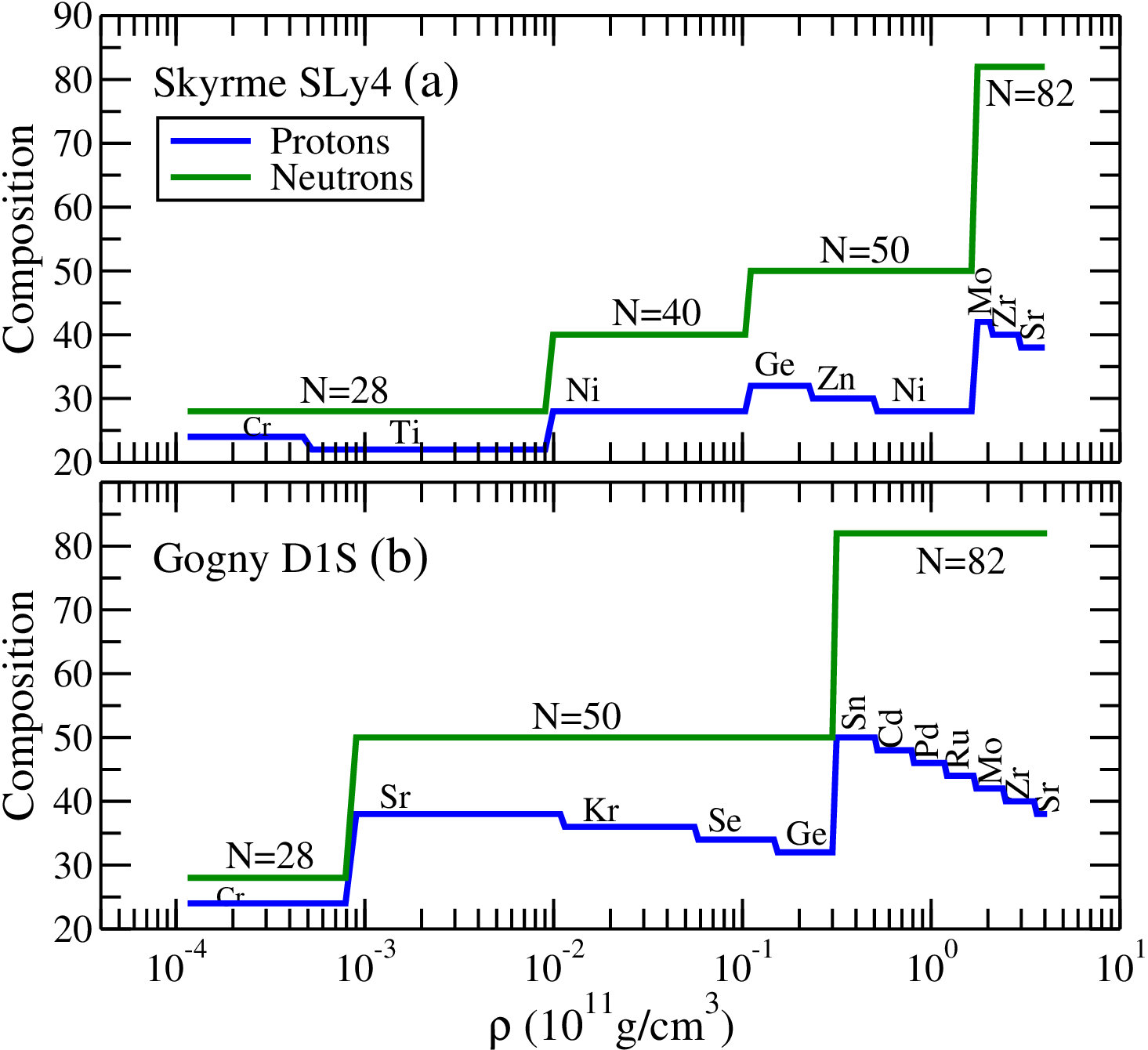}}
 \vspace{-0.2cm}
 \caption{ Composition of the outer crust of a 
           neutron star as predicted by the nonrelativistic mean-field
           models SLy4 (a) and D1S (b). Protons are
           displayed with the (lower) blue line while neutrons 
           with the (upper) green line.}
 \label{Fig09}
\end{figure}

\begin{figure}[t]
  \centerline{ \includegraphics[width=0.6\linewidth]{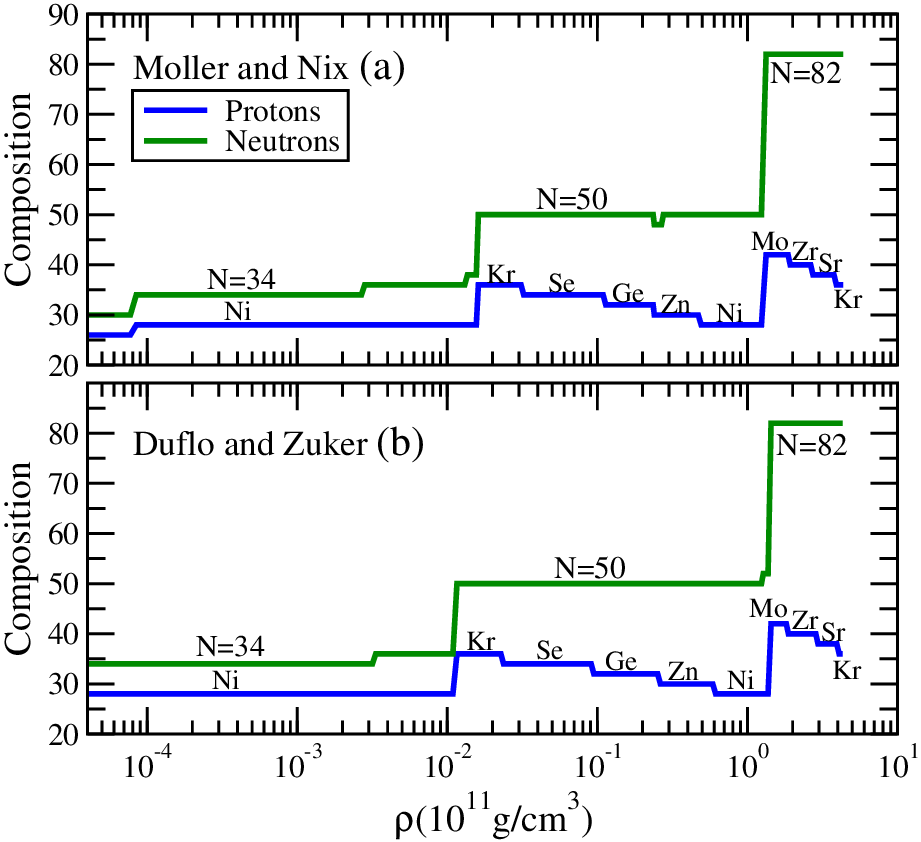}}
 \vspace{-0.2cm}
 \caption{ Composition of the outer crust of a 
           neutron star as predicted using the the mass formulas
           of Moller-Nix (a) and Duflo-Zuker (b). Protons are 
           displayed with the blue (lower) line while neutrons 
           with the green (upper) line.}
 \label{Fig10}
\end{figure}

Similar trends may be observed in Figs.~\ref{Fig08}--\ref{Fig10}
where the composition of the outer crust is displayed as a function of
density. As the system makes a rapid jump in neutron number (say to
magic number $N\!=\!50$) the proton number jumps with it. Along this
neutron plateau, the proton fraction decreases systematically with
increasing density in an effort to reduce the electronic contribution
to the chemical potential. Eventually, the neutron-proton mismatch is
too large and the jump to the next neutron plateau ensues; a jump that
is driven by the symmetry energy. Clearly, the larger the symmetry
energy at low densities, the smaller the neutron-proton mismatch and
the early the jump to the next neutron plateau. These features are
clearly displayed in Fig.~\ref{Fig08} as one contrasts the behavior of
FSUGold to that of NL3. 

Considering Fig.~\ref{Fig08} together with the results predicted by
the nonrelativistic microscopic models SLy4 and D1S, shown in
Fig.~\ref{Fig09}, brings into notice the importance of the nuclear
shell effects in the composition of the outer crust. Indeed, with the
sole exception of $^{58}$Fe ($Z=26$) in the case of the NL3 model at
low crustal density, the ground state of the outer crust consists
systematically of nuclei that have either a neutron magic number
($N=28,50,82$) or a proton magic number ($Z=28,50$), or both.

At the lowest densities of the outer crust, FSUGold and NL3 favor the
presence of the $^{64}$Ni ($Z=28$) and $^{58}$Fe ($Z=26$) isotopes,
respectively. In contrast, at low crustal densities, the
nonrelativistic models SLy4 and D1S favor the major neutron shell
closure $N=28$, allowing nuclei with lower atomic number to populate
the ground state of the crust ---{\it i.e.}, Cr ($Z=24$) in SLy4 and
D1S, and also Ti ($Z=22$) in SLy4.
Compared with D1S, the fact that SLy4 has a lower symmetry energy at
nuclear subsaturation densities (see Fig.~\ref{Fig06}) allows SLy4 to
accomodate a larger neutron-proton mismatch, with the appearance of
the $^{50}$Ti nucleus after $^{52}$Cr. This fact, in combination with
the robustness of the $N=28$ shell, delays in SLy4 the jump to the
next neutron plateau till a much higher crustal density around
$\rho\!=\!2\!\times\!10^{9}{\rm g/cm}^{3}$; the highest density
value found for the first jump to the next plateau in the four
microscopic models studied here. Interestingly enough, SLy4 predicts
that the next neutron plateau occurs at $N=40$ instead of $N=50$,
which is at variance with the other three microscopic models.
Indeed, the possible change of the shell structure and the replacement
of the traditional magic numbers by new islands of stability for
isotopes in the vicinity of the drip lines is a thrilling area of
modern nuclear structure. Our understanding of the outer crust will
benefit from the availability of more information from radioactive
beam experiments on the evolution of the shell structure in exotic
nuclei. Furthermore, some of these exotic isotopes are important also
for other problems in astrophysics such as stellar nucleosynthesis and
element abundances.

Having a similar symmetry energy below nuclear saturation (cf.\
Fig.~\ref{Fig06}), FSUGold and D1S predict a similar composition
for the central plateau of $N=50$. In both models, the $N=50$ plateau
starts with Sr ($Z=38$) and ends with Ge ($Z=32$). The jump to $N=82$
takes place at practically the same crustal density
$\rho\!=\!3\!\times\!10^{10}{\rm g/cm}^{3}$ in FSUGold and D1S\@.
For the $N=82$ plateau, the predicted nuclides are again the same in
these two models, excepting that FSUGold terminates
(namely, $\mu(A,Z;P)\!=\!m_{n}$) at $^{118}$Kr ($Z=36$)
and D1S terminates at $^{120}$Sr ($Z=38$).
We have seen in Fig.~\ref{Fig06} that at nuclear subsaturation
densities the NL3 and SLy4 models have a lower symmetry energy than
FSUGold and D1S\@. Because of this reason, in both NL3 and SLy4 the
plateau of $N=50$ supports more neutron-rich nuclei than FSUGold and
D1S, reaching up to $^{78}$Ni ($Z=28$).
As it happened in the case of FSUGold and D1S, the NL3 and SLy4 
parameter sets jump to the $N=82$ plateau both at a very similar 
crustal density. However, the density value is now higher; {\it i.e.},
$\rho\!\sim\!1.5\!\times\!10^{11}{\rm g/cm}^{3}$, or 5 times larger
than in FSUGold and D1S\@. As expected, the transition to $N=82$ has
been delayed with respect to FSUGold and D1S because NL3 and SLy4 have
a reduced symmetry energy in the subsaturation region. Both NL3 and
SLy4 begin the $N=82$ plateau with Mo ($Z=42$), while NL3 attains the
bottom of the outer crust at $^{120}$Kr and SLy4 at $^{120}$Sr.

In contrast to the microscopic models, fewer differences are
noticeable in Fig.~\ref{Fig10} when comparing the
microscopic/macroscopic model of Moller-Nix to that of Duflo-Zuker.
First, in both models the proton magic nucleus Ni largely dominates
the composition of the ground state of the crust up to a density
$\rho\!\sim\!1-1.5\!\times\!10^{9}{\rm g/cm}^{3}$. This regime is
followed by the $N=50$ and $N=82$ plateaus displaying the same
composition in both models, with only slight differences in the
density value where each isotope occurs. Indeed, in the two
microscopic/macroscopic models the crustal density where the jump to 
$N=82$ ensues is predicted at nearly the same value
$\rho\!\sim\!1.5\!\times\!10^{11}{\rm g/cm}^{3}$. 
This density value was predicted also by the NL3 and SLy4 microscopic
models discussed before.

\begin{figure}[h!]
  \centerline{ \includegraphics[width=0.6\linewidth]{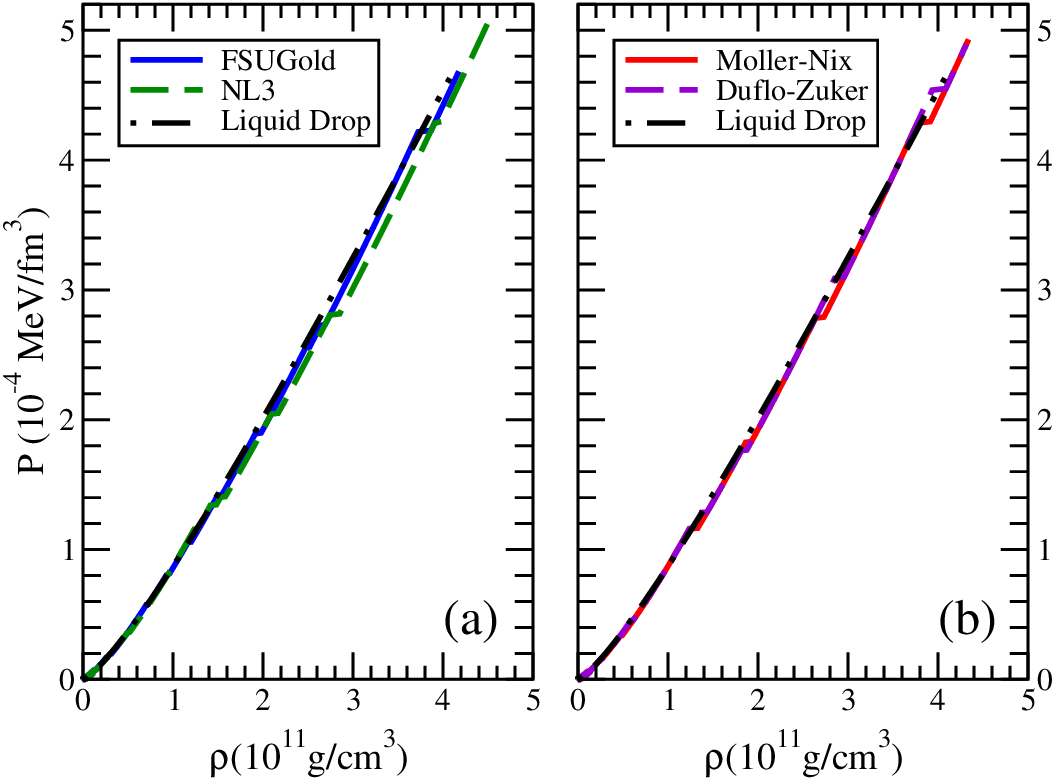}}
 \vspace{-0.2cm}
 \caption{ Panel (a) displays the
          zero-temperature equation of state (pressure {\it vs}
          density) predicted by the accurately calibrated FSUGold 
          (blue solid line) and NL3 (green dashed line) parameter 
          sets. Also shown is the prediction from the simple 
          liquid-drop formula. Panel (b) shows the
          corresponding expression as predicted by the
          Moller-Nix (red solid line) and Duflo-Zuker (purple 
          dashed line) mass formulas.}
 \label{Fig11}
\end{figure}

We conclude this section by displaying in Fig.~\ref{Fig11} the predictions for the
equation-of-state (pressure {\it vs} density relation) of the
outer crust of a neutron star. The left-hand panel shows results from
calculations using the FSUGold (blue solid line) and NL3 (green dashed
line) parameter sets. (The results obtained with D1S and SLy4 display
the same general trends and therefore are not shown.)
Although barely visible in the figure, the density shows
discontinuities at those places where the composition changes
abruptly. It is also noted that the FSUGold parametrization predicts a
pressure that rises slightly faster with density than NL3. For the NL3
set, the symmetry energy admits lower values of the proton/electron
fraction $y$ which, in turn, lowers the pressure of the system. Lower
values of $y$ also yield lower values of the chemical potential,
thereby delaying the arrival to the neutron-drip line. Indeed, 
whereas FSUGold predicts a drip-line density of 
$\rho_{\rm drip}\!=\!4.17\!\times\!10^{11}{\rm g/cm}^{3}$, with
NL3 the transition is delayed by about 8\%, or until
$\rho_{\rm drip}\!=\!4.49\!\times\!10^{11}{\rm g/cm}^{3}$.
A similar plot is shown for the microscopic/macroscopic models of
Moller-Nix (red solid line) and Duflo-Zuker (purple dashed line).
Differences among these two models are barely noticeable. Indeed,
drip-line densities in both models are predicted at about $\rho_{\rm
drip}\!=\!4.3\!\times\!10^{11}{\rm g/cm}^{3}$. Model predictions for
various observables at the base of the outer crust ({\it i.e.}, in the
drip-line region) are listed in Table~\ref{Table1} for all the
studied models.

\begin{table}[t]
\begin{center}
\begin{tabular}{|r|cccccc|}
 \hline
 Model  &  $\rho$ &   $ n$ 
        &  $P$ &   $\mu_{e}$   &  Element 
        &  $B/A$ \\ 
 \hline
 Moller-Nix   &  4.34  &  2.60  &  4.93  &  26.22  &  $^{118}$Kr &   $7.21$ \\
 Duflo-Zuker  &  4.32  &  2.58  &  4.89  &  26.17  &  $^{118}$Kr &   $7.19$ \\
 FSUGold      &  4.17  &  2.50  &  4.68  &  25.88  &  $^{118}$Kr &   $7.11$ \\
 NL3          &  4.49  &  2.69  &  5.06  &  26.39  &  $^{120}$Kr &   $7.13$ \\
 SLy4         &  4.10  &  2.46  &  4.81  &  26.06  &  $^{120}$Sr &   $7.42$ \\
 D1S          &  3.98  &  2.39  &  4.62  &  25.81  &  $^{120}$Sr &   $7.34$ \\
\hline
\end{tabular}
\end{center}
\caption{Equation-of-state observables
  (mass density $\rho$ in $10^{11}{\rm g/cm}^{3}$,
  baryon density $n$ in $10^{-4}{\rm fm}^{-3}$,
  pressure $P$ in $10^{-4}{\rm MeV/fm}^{3}$,
  and electronic chemical potential $\mu_{e}$ in MeV)
  and the predicted composition
  (nucleus and binding-energy per nucleon $B/A$ in MeV)
  at the base of the outer crust.}
\label{Table1}
\end{table}

%
%
\section{Conclusions, final remarks and open questions}
\label{Conclusions}

We studied the structure and composition of the outer crust of a 
nonaccreting neutron star focusing on the effects of the current 
uncertainties in the nuclear symmetry energy and its density derivative 
on the crustal properties.
For that, we followed the seminal work by Baym, Pethick, and Sutherland, as 
well as the more recent comprehensive work by Ruester, Hempel, and
Schaffner-Bielich. In our investigations, six different models
were adopted. Two of these models, Moller-Nix and Duflo-Zuker, are
based on a combined microscopic/macroscopic approach and yield the
most accurate nuclear masses available in the literature. 
The other four models are of a purely microscopic nature. They are based
on relativistic (NL3 and FSUGold) and nonrelativistic (Skyrme SLy4 and
Gogny D1S) mean-field approaches. Although the microscopic/macroscopic
models are significantly more accurate than the mean-field models in
the description of nuclear masses, microscopic models have the
advantage of making definite predictions on how the symmetry energy
changes with density (see Figs.~\ref{Fig05} and~\ref{Fig06}).

The composition and equation of state of the outer crust emerge from a
competition among the nuclear, electronic, and lattice contributions to 
the energy, pressure and chemical potential of the system.
The nuclear contribution 
is independent of the density of the system and, therefore, does not contribute 
to the total pressure; it appears exclusively in the form of nuclear masses.
The electronic contribution is modeled as a zero-temperature 
free Fermi gas and dominates the behavior of the system with baryon density. Finally, 
the lattice contribution which also depends on the density, provides a modest 
correction, less than a 10\%, to the total energy of the system. Hence, the 
competition between the different terms is basically driven by the energy of the 
electron Fermi gas and the nuclear symmetry energy. The former favors a small electron 
fraction and, to preserve charge neutrality, a small proton fraction which 
leads the system toward more and more neutron-rich nuclei in the crust. The 
nuclear symmetry energy opposes such a change by driving the system to
a more symmetric configuration in terms of the neutron and proton
numbers. 

For a better understanding of the competition between the electron
contribution and the nuclear symmetry energy, we
implemented a ``toy model'' of the outer crust by using a simple
semi-empirical ({\sl ``Bethe-Weizs\"acker''}) nuclear mass
formula. Volume, surface, Coulomb, and asymmetry terms were extracted
from a least-squares fit to 2049 nuclei. Such a simple model provided 
us with useful insights thanks to the analytic structure of the
results. Indeed, a particularly transparent result was obtained
that illustrates nicely the role of the electronic contribution and the
nuclear symmetry energy in deciding the proton fraction:
\begin{equation}
 y(p^{}_{\rm F})=y_{0}-\frac{p^{}_{{\rm F}e}}{8a_{\rm a}}
               +{\mathcal O}(p^{2}_{{\rm F}e})\;,
\end{equation}
where $y_{0}$ is the zero-density proton fraction, $p^{}_{{\rm F}e}$ is the 
electronic Fermi momentum and $a_{\rm a}$ is the symmetry energy coefficient
of the liquid-drop mass formula.
While illuminating, this (first-order) result is also surprisingly accurate, 
as the electronic Fermi momentum at the base of the outer crust is very 
close in value to the symmetry energy coefficient 
($p^{}_{{\rm F}e}\!\approx\!26$~{\rm MeV} {\it vs} $a_{\rm a}\!\approx\!23$~{\rm MeV}). 
In particular, the toy model predicts a value for the electron fraction 
at the base of the crust that differs by only a few percent from that of
the drip-line nucleus ${}^{118}$Kr.

How the symmetry energy coefficient $a_{\rm a}$ is modified as nuclei
move far away from the line of stability is not known. Likely, the
symmetry energy is reduced in neutron-drip nuclei due to the
development of a dilute neutron skin. Mean field models of nuclear
structure predict that the size of the neutron skin of neutron-rich
nuclei is strongly correlated with the slope $L$ of the symmetry
energy. The stiffer the symmetry energy, i.e., the larger the $L$
parameter, the thicker the neutron skin.
To investigate the sensitivity of the structure and composition of the outer crust to 
the density dependence of the symmetry energy, we employed relativistic
(NL3 and FSUGold) and nonrelativistic (SLy4 and D1S) mean-field models.
Although these models have been accurately calibrated for the description 
of masses, charge radii, and other important ground-state properties,
they predict a significantly different density dependence for the symmetry
energy. Most of the relativistic models predict larger values of the symmetry energy at saturation than
the nonrelativistic ones. It is also quite common that those mean-field models
show an opposite trend at subsaturation densities. 
That is, those models with a stiffer symmetry energy at saturation
predict smaller values of the symmetry energy at subsaturation
densities (see Figs.~\ref{Fig05} and~\ref{Fig06}). One of the main
goals of the present chapter was to document how such differences
impact on the composition of the outer crust.  

Quite generally, we show that the first substantial change to the ground-state 
configuration of the nuclei composing the lattice at very low densities corresponds 
to a jump of the neutron number that remains fixed at the magic number
$N\!=\!50$ for a wide range of densities. At the same densities, 
the proton fraction also suffers a jump and, then, decreases systematically in an 
effort to reduce the electronic contribution to the chemical potential. Eventually, 
the proton fraction becomes too low and the penalty caused by the symmetry 
energy at subsaturation densities drives the system into the next plateau at the magic number $N\!=\!82$ which 
remains fixed until the outer-inner crust interface is reached. How low can the proton fraction get and,
consequently, how exotic is the composition of the crust is, then, a
question that must be answered by the symmetry energy: its accurate determination remains 
as one of the most outstanding problems in Nuclear Physics nowadays. Indeed, whereas
NL3 predicts the formation of ${}^{78}_{28}$Ni${}^{}_{50}$, FSUGold
(having a larger symmetry energy) leaves the $N\!=\!50$ plateau with
the formation of ${}^{82}_{32}$Ge${}^{}_{50}$ (or four protons
earlier); similar conclusions could be drawn from the study of the
nonrelativistic models SLy4 and D1S.
This result may be stated in the form of a correlation
between the neutron radius of\/ ${}^{208}$Pb and the composition of the
outer crust: {\emph{the larger the neutron skin of ${}^{208}${\rm Pb},
the more exotic the composition of the outer crust}}. 
Finally, and as it was done in Ref.~\cite{Ruester:2005fm}, we have
computed crustal properties using two of the most accurate tables of
nuclear masses available today, namely, those of Moller-Nix and
Duflo-Zuker. Our results using the model of Moller and Nix agree well
with those published in Ref.~\cite{Ruester:2005fm}. These results are
practically indistinguishable from the ones obtained using the
Duflo-Zuker nuclear mass table; a table that includes 9210 nuclei! Both 
calculations served us as a reference and test of the formalism used for 
our study of the outer crust of a nonaccreting neutron star. 

With the advent of new rare ion-beam facilities, the experimental
study of exotic nuclei will be increasingly feasible in laboratories
worldwide. This promising scenario, together with the recent and,
hopefully, future observations of crustal modes in magnetars, and
perhaps the detection of gravitational waves from oscillating neutron
stars, will likely provide us with more stringent limits on the
equation of state of neutron-rich matter and will contribute to a
better understanding of the outer crust of neutron stars.

%
%
\section*{Acknowledgments}
M.C. partially supported by the Spanish Consolider-Ingenio 2010
Programme CPAN CSD2007-00042 and by Grants No.\ FIS2008-01661 from
MICINN (Spain) and FEDER and No.\ 2009SGR-1289 from Generalitat de
Catalunya (Spain). J.P. was supported in part by DOE grant 
DE-FG05-92ER40750.

%
%

\label{lastpage-01}

\end{document}